\documentclass{aa}  

\usepackage{graphicx}
\usepackage[dvipsnames]{xcolor}
\usepackage{txfonts}
\usepackage[colorlinks=true,
    linkcolor=blue,
    citecolor=blue,
    filecolor=magenta,      
    urlcolor=cyan]{hyperref}
\usepackage{ulem}
\usepackage{orcidlink}

\begin{document}

   \title{Emission signatures from sub-pc Post-Newtonian binaries embedded in circumbinary discs}

   \titlerunning{PN evolution of MBHBs in discs}
   \authorrunning{A. Franchini et al.}

   \author{Alessia Franchini\orcidlink{0000-0002-8400-0969}
          \inst{1,}
          \inst{2,}
          \inst{3}\fnmsep\thanks{alessia.franchini@unimib.it}
          \and
          Matteo Bonetti\orcidlink{0000-0001-7889-6810}
          \inst{1,}
          \inst{3,}
          \inst{4}
          \and
          Alessandro Lupi\orcidlink{0000-0001-6106-7821}
          \inst{5,1,3}
          \and
          Alberto Sesana\orcidlink{0000-0003-4961-1606}
          \inst{1,}
          \inst{3}
    }

   \institute{
            Dipartimento di Fisica ``G. Occhialini'', Universit\`a degli Studi di Milano-Bicocca, Piazza della Scienza 3, I-20126 Milano, Italy
        \and
            Center for Theoretical Astrophysics and Cosmology, Institute for Computational Science, University of Zurich,
Winterthurerstrasse 190, CH-8057 Zürich, Switzerland
        \and
            INFN, Sezione di Milano-Bicocca, Piazza della Scienza 3, I-20126 Milano, Italy
        \and
            INAF - Osservatorio Astronomico di Brera, via Brera 20, I-20121 Milano, Italy
        \and
            DiSAT, Universit\`a degli Studi dell'Insubria, via Valleggio 11, I-22100 Como, Italy
      }

   \date{Received ; accepted }

 
  \abstract{
  We study the dynamical evolution of quasi-circular equal mass massive black hole binaries  embedded in circumbinary discs from separations of $\sim 100R_{\rm g}$ down to the merger, following the post merger evolution. The binary orbit evolves owing to the presence of the gaseous disc and the addition of Post-Newtonian (PN) corrections up to the 2.5 PN order, therefore including the dissipative gravitational wave back-reaction. We investigate two cases of a relatively cold and warm circumbinary discs, with aspect ratios $H/R=0.03,\,0.1$ respectively, employing 3D hyper-Lagrangian resolution simulations with the {\sc gizmo}-MFM code. We extract spectral energy distributions and light curves in different frequency bands (i.e. X-ray, optical and UV) from the simulations. We find a clear two orders of magnitude drop in the X-ray flux right before merger if the disc is warm while we identify a significant increase in the UV flux regardless of the disc temperature.
  The optical flux shows clear distinctive modulations on the binary orbital period and on the cavity edge period, regardless of the disc temperature.  
  We find that the presence of a cold disc can accelerate the coalescence of the binary by up to 130 seconds over the last five days of inspiral, implying a phase shift accumulation of about $\pi\,$radians compared to the binary evolution in vacuum. These differences are triggered by the presence of the gaseous disc and might have implications on the waveforms that can be in principle detected. 
  We discuss the implications that these distinctive signatures might have for existing and upcoming time domain surveys and for multimessenger astronomy.  }

   \keywords{galaxies:active -- galaxies:nuclei -- quasars:supermassive black holes -- Black hole physics -- Relativistic processes }

   \maketitle
%
\section{Introduction}

The vast majority of massive galaxies is known to host a Massive Black Hole (MBH) at their centre \citep[][and references therein]{Kormendy2013}. When two galaxies merge, dynamical friction causes the two hosted MBHs to sink to the centre of the merger remnant where they form a bound binary \citep{begelman1980}.
Massive black hole binaries (MBHBs) then further evolve towards coalescence through interaction with the environment, i.e. stars and gas \citep{ArmitageNatarajan2002,Dotti2007,Lodato2009}, before entering the regime where further orbital evolution is dominated by gravitational wave (GW) emission \citep{Peters1964}.

The evolution of the binary through the interaction with stars has been investigated by several authors, and has been found to efficiently shrink the binary to the GW emission stage and eventually to coalescence \citep{Khan2011,Vasiliev2014,Gualandris2017,Bortolas2021,Varisco2021}. In gas rich environments, however, the sub-parsec evolution of a MBHB can be dominated by its interaction with a circumbinary accretion disc \citep{escala2005,Dotti2007,Cuadra2009}. 

The interaction between a binary and its circumbinary disc has been recently extensively investigated and the outcome strongly depends on the disc properties. Numerical studies employing 2D fixed binary orbit calculations have showed that warm discs can drive the binary apart through the torques exerted by the disc material that leaks into the cavity carved by the binary into the disc \citep{Duffell2019,Munoz2020,tiede2020}. However, it has been also shown, mainly through 3D live binary orbit calculations \citep{ragusa2016,heathnixon2020,franchini2021,Franchini2022,Franchini2023}, that when the disc is colder (and thinner), the central MBHB is much more effective in preventing material to enter the cavity, resulting in net binary hardening.
Those studies employed scale free simulations as they did not include any scale-dependent process, such as the evolution of the binary driven by GWs emission, which dominates the evolution towards the merger once the binary decouples from the disc \citep{ArmitageNatarajan2002}.
The investigation of the gaseous dynamics in the GW-driven regime is fundamental in order to unveil the possible electromagnetic (EM) counterparts of the merger \citep{Bogdanovic2022}.

When the two MBHs are at such close separations, \cite{2017ApJ...838...42B} found that the quasi-periodic exchange of  material between the two circumBH discs (also called mini-discs), together with their periodic mass fluctuations caused by their inability to maintain equilibrium inflow, possibly provides an identifiable source of  variability in their EM emission. 
\cite{roedig2014} investigated instead the hotspots generated by shocks between the gaseous stream that feeds the circumBH discs onto the discs themselves, finding however that the total energy release in such shocks might not be sufficiently large for it to be detected.
A number of studies in the literature have investigated the thermal spectrum from the circumbinary and circumBH discs and the imprint in it created by the cavity carved by the binary \citep{2012MNRAS.425.2974T,2018MNRAS.476.2249T,2018ApJ...865..140D}.
In particular, \cite{2012MNRAS.425.2974T} showed that at separations of the order of $20$ gravitational radii, the contribution from the circumBH discs is only 1\% that of the circumbinary disc.
\cite{2018ApJ...865..140D} investigated the emission on the same small scales assuming that circumBH discs, the streams and the circumbinary disc are sources of both thermal and coronal emission. Their model produces thermal radiation with a spectrum that differs only modestly from ordinary single MBH systems.

A proper characterization of possibly distinctive EM signals coming from these systems is of paramount importance for the identification of counterparts to the MBHBs that will be targeted by the Laser Interferometer Space Antenna \citep[LISA,][]{2023LRR....26....2A}. Moreover, the EM identification of extremely massive MBHBs might aid the interpretation of the GW signal recently detected by pulsar timing array (PTA) collaborations around the globe; namely the EPTA \citep{2023arXiv230616214A,2023arXiv230616224A,2023arXiv230616225A,2023arXiv230616226A,2023arXiv230616227A,2023arXiv230616228S}, NANOGrav \citep{nanograv2023,2023ApJ...951L...8A,2023ApJ...951L...9A,2023ApJ...951L..10A,2023ApJ...951L..11A},  PPTA \citep{ppta2023} and CPTA \citep{2023RAA....23g5024X}.

A recent further attempt in this direction has been made by two numerical studies that have investigated the dynamics of the gas that surrounds a MBHB at $\sim 100$ gravitational radii separation, evolving by means of GWs only \citep{Dittmann2023,MajorKrauth2023}.
In particular, \cite{MajorKrauth2023} found a persistent thermal X-ray emission lasting until 1-2 days before the merger and then abruptly dropping by several orders of magnitude, which is consistent with the complete accretion of the material orbiting around each binary component (i.e. the mini-discs).
\cite{Dittmann2023} used a different numerical code and found the decoupling to occur in the LISA band only for very high disc viscosities (i.e. kinematic viscosity $\nu \geq 0.03$). 
These studies employ 2D hydrodynamics simulations where the binary orbit evolves only through GW emission according to an analytic prescription, without integrating the binary orbit over time. This implies that the modification to the binary orbit due to its interaction with the gas, which dominates the evolution before the decoupling, is neglected.

In this work, we make a step further by investigating the interaction between a gaseous circumbinary disc and an equal mass binary whose dynamics is described with Post-Newtonian (PN) correction up to the 2.5 PN order. This approach allows us to accurately assess possible EM signatures of merging MBHBs in a regime where the binary naturally evolves as a result of the interaction with both the gaseous circumbinary disc and via GWs emission. 
The main aim of our work is to characterize the thermal emission coming from the gaseous disc and its changes as the binary evolves towards the merger and to identify 
possible EM precursors or signatures of merging MBHBs. 

The paper is organized as follows. In Section \ref{sec:setup} we describe our numerical setup. We then present our results in Section \ref{sec:results} and discuss their implications while we draw our conclusions in Section \ref{sec:conclusions}.

\section{Numerical Setup}
\label{sec:setup}

The initial conditions for this work consist of a binary surrounded by a quasi-steady state circumbinary gaseous disc. 
Specifically, we take the 1000th orbit of two Newtonian simulations we performed in \cite{Franchini2022} with disc aspect ratio $H/R=0.03,\,0.1$ and we then follow the evolution until merger applying PN corrections to the binary dynamics as outlined in Section \ref{sec:pn}. 

The initial number of particles is different between the two simulations as the gas dynamics strongly depends on the disc aspect ratio, i.e. on its temperature. 
The cold disc simulation ($H/R=0.03$) has roughly $4\times 10^6$ particles, while the warm disc is modelled using $2\times 10^6$ particles.\footnote{ We note that since we have performed convergence tests in our previous work \citep{Franchini2022}, we do not repeat those tests with PN corrections as we have reached a high enough resolution for the torques value to converge.} The number of particles is set by previous simulations \citep{Franchini2022}, where the initial number of particles was the same, regardless of the choice of disc aspect ratio, and the splitting scheme was identical. In this respect, the lower number of particles at the 1000th orbit of the warm disc simulation is simply due to the higher viscosity, which results in significantly more material entering the binary cavity and accreting on the two MBHs compared to the cold disc case. 

The circumbinary discs initially extended from $R_{\rm in} = 2a$ to $R_{\rm out} = 10a$. The radial extension is however slightly modified by the interaction with the binary over the first 1000 orbits, resulting in a cavity radius around $3a$ (slightly larger in the colder disc simulation). The gas is described by a locally isothermal equation of state with the sound speed $c_{\rm s}$ defined by Eq.~(4) in \cite{farris2014}, and an initial surface density profile scaling as $\Sigma \propto R^{-3/2}$, normalised to get a total mass $M_{\rm disc}=100M_{\odot}$ in both simulations. We here also include the effect of gas viscosity entering the Navier-Stokes fluid equations as described in \citet{hopkins2016}, assuming a shear viscosity in the disc $\nu=\alpha c_{\rm s}H$, parametrised using a viscosity parameter $\alpha=0.1$ \citep{SS1973}, and no bulk viscosity. 

The equal mass binary has an initial mass $M=M_1+M_2=10^6M_{\odot}$ and we choose the initial separation to be twice the expected decoupling radius \citep{ArmitageNatarajan2002}. Below this separation, the dynamics of the binary is governed by GW emission, while at larger distances the gaseous interaction is predominant with respect to GWs. 
Assuming the viscosity coefficient $\alpha=0.1$, the decoupling radius for $H/R=0.03$ is $56\,R_{\rm g}$ while for $H/R=0.1$ is $22\,R_{\rm g}$, where $R_{\rm g}=GM/c^2$ \citep{ArmitageNatarajan2002}. 
We therefore start the cold and warm disc simulations with separations $a=120\,R_{\rm g}$ and $a=60\,R_{\rm g}$ respectively. 
The initial binary orbital period $P_{\rm b}=2\pi(a^3/GM)^{1/2}$ is therefore $0.47$ days for the cold disc and $0.17$ days for the warm one. 
This corresponds to an initial GW frequency of $5\times10^{-5}$Hz and $1.4\times10^{-4}$Hz respectively, which is at the lower end of the frequency range covered by LISA. 

For the simulations, we employ the code {\sc gizmo} \citep{Hopkins2015} in its 3D mesh-less finite mass (MFM) mode.
In order to increase the resolution when and where necessary, without globally slowing down our simulations, we employ an on-the-fly adaptive particle-splitting approach, which is similar in spirit to the adaptive mesh refinement of grid-based codes. Such technique represents a natural generalisation of the refinement/de-refinement scheme exploited by {\sc arepo} \citep{Springel2010} and {\sc gizmo} \citep{Hopkins2015} to maintain an almost constant mass per cell during a simulation in a finite-volume scheme.
In this work, we exploit the particle splitting algorithm in {\sc gizmo} to split gas particles that enter a sphere of radius $r_{\rm ref}=4a$ centred on the centre of mass of the binary-disc system. The refinement scheme is the same as the one adopted in \cite{Franchini2022} with a maximum refinement of a factor $32$. 

Every time a gas particle approaches one of the sinks by entering its sink radius $r_{\rm sink} = 0.05a$, we flag it as eligible to be accreted. 
As the binary orbit is allowed to vary \citep{Franchini2023}, conservation of mass and linear and angular momentum are ensured during each accretion event, in the same way it is done in the {\sc phantom} code \citep{bate1995}.

In this work, we neglect the disc self-gravity (see \citealt{franchini2021} for self-gravity treatment) and use the version of the code described in details in \cite{Franchini2022}.  The reason for this choice is that the inclusion of the disc self-gravity would make the evolution more complex, as other processes would come into play (e.g. cooling, gravitational instabilities), and we wanted to investigate the effect of PN corrections on the gas dynamics in the simplest possible scenario. In addition, self-gravity can be neglected for sufficiently compact binaries, for which the disc mass is negligible
(see \citealt{franchini2021} for a discussion). The extension of this work to the self-gravity regime will be performed in a future study. 

We highlight the fact that we evolve the binary orbit \citep{Franchini2023} considering both the interaction with the gas and the PN corrections to the binary dynamics.

\begin{figure}
    \centering
    \includegraphics[width=\columnwidth]{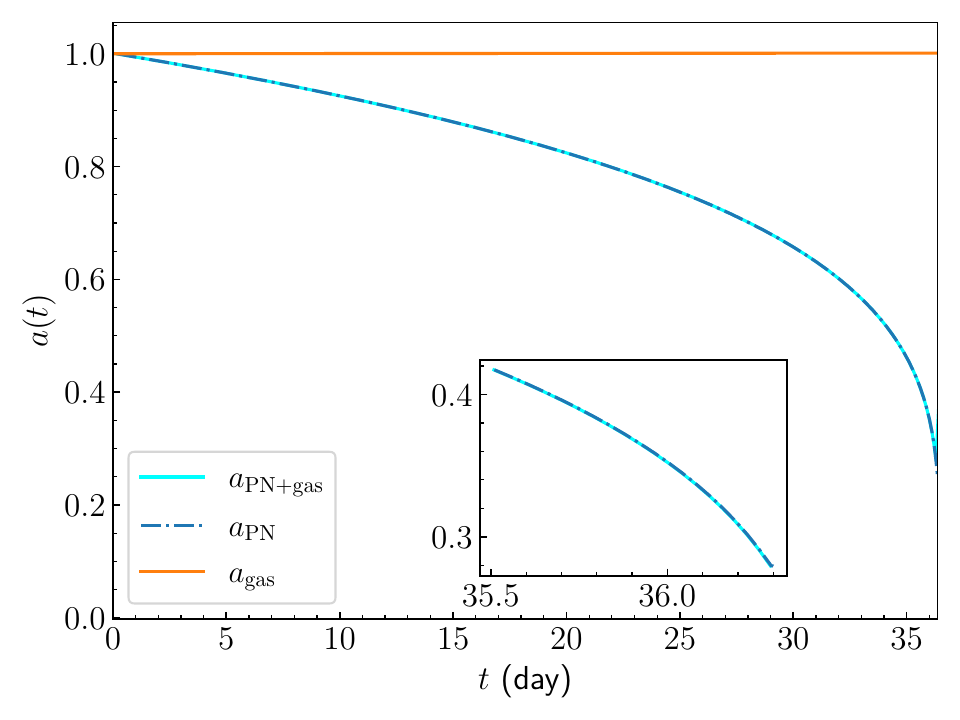}
    \includegraphics[width=\columnwidth]{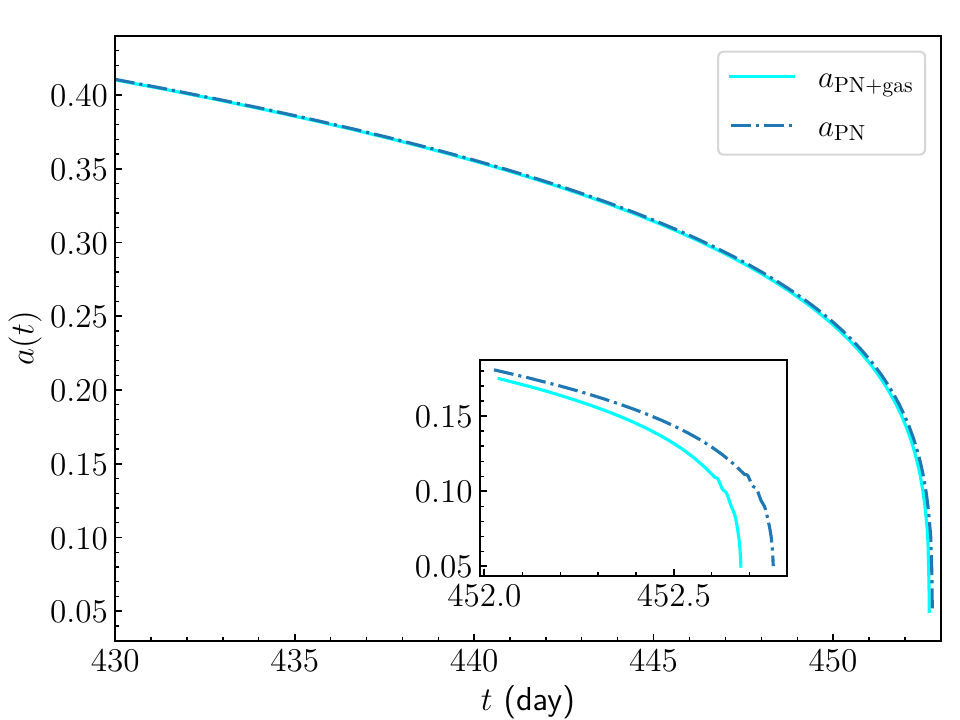}
    \caption{Evolution of the binary semi-major axis. The orange line shows the evolution caused by the mere interaction with the gaseous disc with $H/R=0.1$ in the Newtonian simulation (see right panel of Fig. 2 in \cite{Franchini2022}). The cyan line shows the evolution including both the PN corrections and the gaseous torque while the blue dot-dashed line shows the evolution caused by PN corrections only. The inset shows the deviation due to the gas torque on the binary. The upper panel refers to the warm disc simulation, where the gas is expanding the binary, while the lower panel shows the cold disc case, where the binary inspiral is aided by the gaseous disc. }
    \label{fig:avst}
\end{figure}

\begin{figure}
    \centering
    \includegraphics[width=\columnwidth]{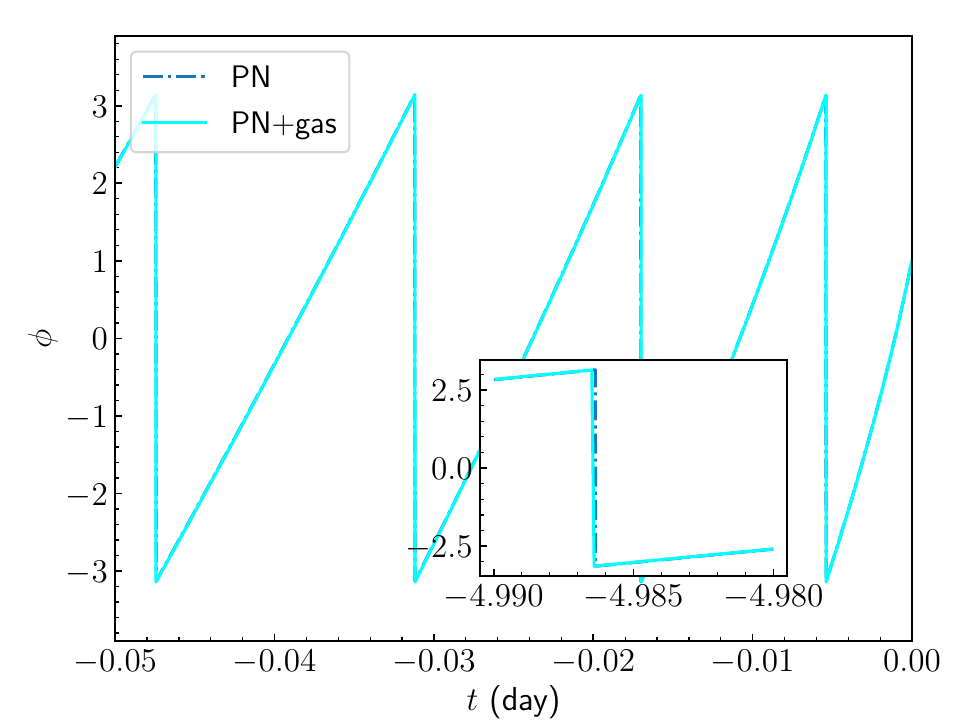}
    \includegraphics[width=\columnwidth]{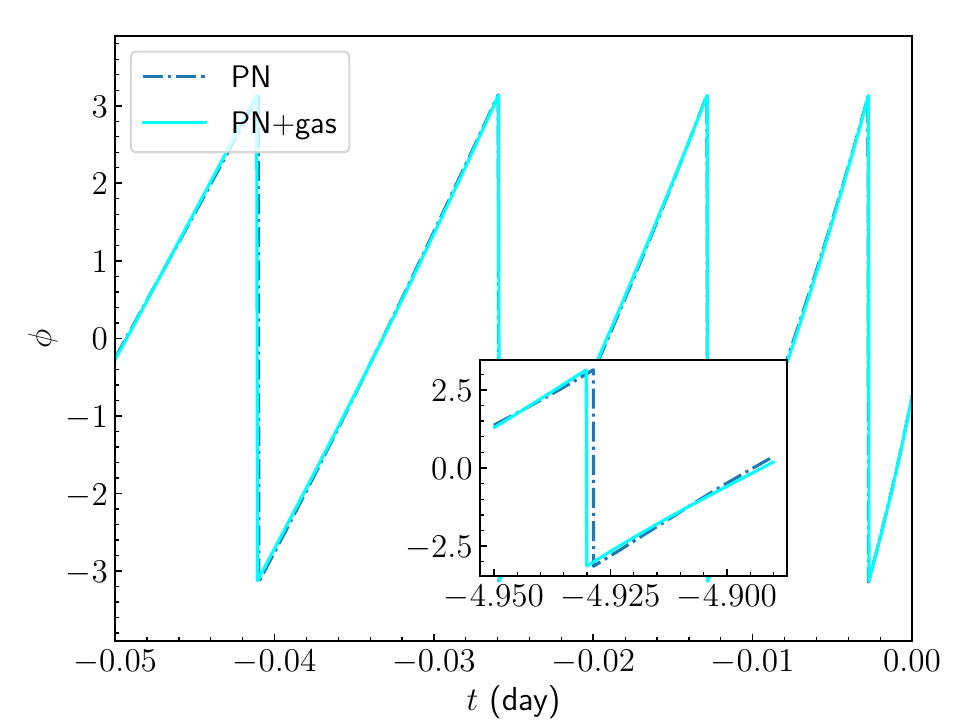}
    \caption{Orbital phase shift between the run with the warm (upper panel) and cold (lower panel) gaseous disc contribution to the binary evolution (cyan line) and the one where we only integrate the PN binary orbit (blue dot-dashed line). }
    \label{fig:phase}
\end{figure}

\begin{figure*}
    \centering
    \includegraphics[width=\textwidth]{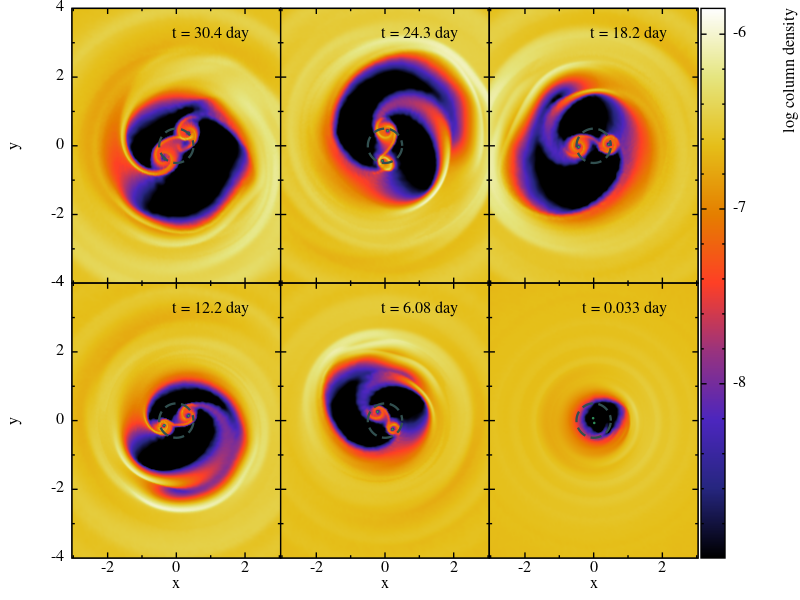}
    \caption{Column density maps of the warm disc simulations. The disc has aspect ratio $H/R=0.1$. The 6 panels show the system starting from $30.4$ days prior to the merger. The dashed green circle shows the initial binary orbit.}
    \label{fig:warm}
\end{figure*}

\begin{figure*}
    \centering
    \includegraphics[width=\textwidth]{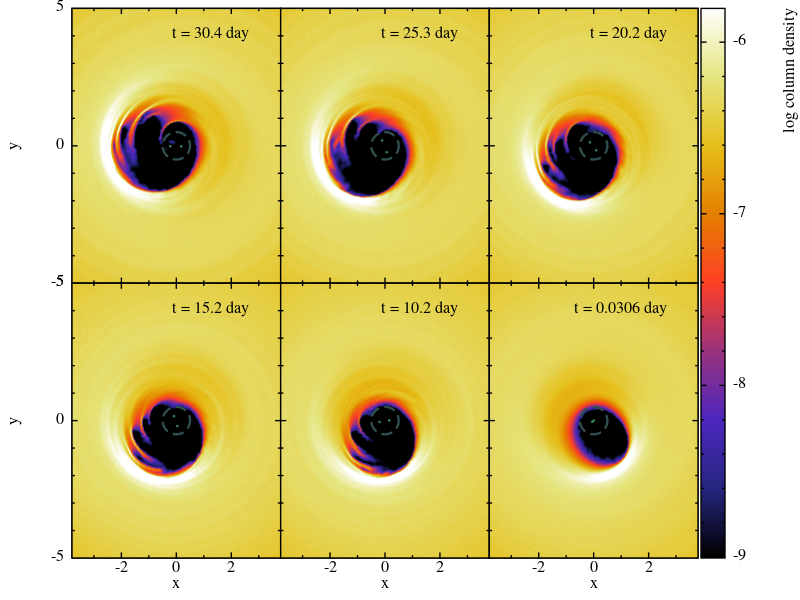}
    \caption{Column density maps of the cold disc simulations. The disc has aspect ratio $H/R=0.03$. The 6 panels show the system starting from $30.4$ days prior to the merger. The dashed green circle shows the initial binary orbit.}
    \label{fig:cold}
\end{figure*}

\subsection{Post-Newtonian corrections}
\label{sec:pn}

We integrate the equations of motion of the binary system considering PN corrections up to 2.5PN order. This includes the conservative 1PN and 2PN terms as well as the GW emission encoded into the 2.5PN term.
We assume the MBHs that form the binary to be non-spinning. We defer the investigation of spin effects to a future work.
We consider the equations of motion derived using the Lagrangian formalism of \citet{2014LRR....17....2B}. Practically, we evolve forward in time the positions and velocities of the two particles with mass $m_1, m_2$ according to the PN accelerations only retaining terms up to $1/c^5$, i.e.:
\begin{align}
    \frac{d^2 \mathbf{r}_1}{dt^2} &= -\frac{G m_2}{r^2}\mathbf{n}_{12} \nonumber\\
    & +\frac{1}{c^2}\biggl( \mathcal{A}_{\rm 1PN} \ \mathbf{n}_{12} + \mathcal{B}_{\rm 1PN} \mathbf{v}_{12} \biggr) \nonumber\\
    &+\frac{1}{c^4}\biggl( \mathcal{A}_{\rm 2PN} \ \mathbf{n}_{12} + \mathcal{B}_{\rm 2PN} \mathbf{v}_{12} \biggr) \nonumber\\
    & + \frac{1}{c^5}\biggl( \mathcal{A}_{\rm 2.5PN} \ \mathbf{n}_{12} + \mathcal{B}_{\rm 2.5PN} \mathbf{v}_{12} \biggr)
    + \mathcal{O}\left(\frac{1}{c^6}\right), 
\end{align}
where $\mathbf{n}_{12} = (\mathbf{r}_1-\mathbf{r}_2)/|\mathbf{r}_1-\mathbf{r}_2|$, $\mathbf{v}_{12} = (\mathbf{v}_1-\mathbf{v}_2)$, while $r = |\mathbf{r}_1-\mathbf{r}_2|$. We defer the reader to Eq. 203 of \citet{2014LRR....17....2B} for the expressions of the coefficients $\mathcal{A}_{\rm xPN}$ and $\mathcal{B}_{\rm xPN}$, with the convention that the acceleration for the particle 2 can be readily obtained by exchanging all the particle labels $1 \leftrightarrow 2$. 

The dependence of the above accelerations on the particle velocities complicates the numerical time evolution of the system, implying that a standard leap-frog integration algorithm (``kick-drift-kick'') is not well suited for integrating the PN corrections. We therefore implement an intermediate predictor step to update the particle velocities at the end of the time-step, accounting for the PN corrections, re-enforcing the numerical stability of the integration algorithm. Our approach is similar to the one outlined in Section 6.2 of \cite{Liptai2019}, except that we use a predictor-corrector approach instead of implementing the implicit ``kick-drift-kick''. 
We tested our numerical integration by running a simulation without the gaseous disc and comparing the merger timescale with the theoretical expectation, finding almost perfect agreement.

\subsection{Spectral energy distribution}

Since we are ultimately interested in understanding the potentially observable binary features in the EM emission, we compute the spectral energy distribution (SED) and the luminosity in different bands assuming black-body emission from each resolution element.

Although we do not explicitly include radiation pressure in our simulations, as this is quite costly, we do assume both discs to be radiation pressure dominated and therefore calculate the disc temperature from the gas pressure according to 
\begin{equation}
    P = \frac{4\sigma_{\rm SB}}{3c} T_{\rm c}^4
\end{equation}
where $\sigma_{\rm SB}$ is the Stefan-Boltzmann constant and $T_{\rm c}$ is the disc mid-plane temperature. We calculated $T_{\rm c}$ by averaging the temperature of each particle within $0.5\,H/R$ from the midplane (i.e. $z=0$), weighting it on the local density $\rho$. 
The effective disc temperature is related with the mid-plane temperature through

\begin{equation}
    T_{\rm eff}^4 = \frac{4}{3}\frac{T_{\rm c}^4}{\kappa \Sigma}
\end{equation}
where $\kappa = 0.4 \,{\rm cm^2 g^{-1}}$ is the electron scattering opacity and $\Sigma$ is the disc surface density.
We assume that only the optically thick (i.e. those with $\tau = \kappa \Sigma \geq 1$) regions of the domain emit thermal black body radiation.
We note that we do not compute the emission coming from regions where our numerical resolution is low, i.e. we set a minimum threshold for the surface density. The main reason is that there is spurious, non physical emission coming from the boundary between the particles kernel and the cavity in both the circumBH disc outer edges and the cavity wall.
This might have some impact on the magnitude of the X-ray flux but we can confidently capture variations of the X-ray flux due to the gas dynamics during the inspiral.

We assume the observer to be looking the disc face-on, neglecting any Doppler effect. 
We extract temperature maps from the simulation snapshots and divide the $x-y$ plane in a grid of 800$\times$800 pixels. We then calculate the luminosity produced at each frequency within the range $10^{14}-10^{19}$ Hz by dividing the simulation domain into three regions: the circumBH discs region $r<a$, the stream region $a<r<3a$ and the circumbinary disc $r>3a$. Note that as $a$ evolves the contribution to the spectrum coming from these regions changes in time.

We then produce light curves in different bands by integrating the SED over the optical ($1.8-3.1$ eV), UV ($3.1-124.0$ eV) and X-ray ($124.0$ eV $-124$ keV) band.
We finally compute the luminosity emitted by an element with a surface $dS$ in each frequency band as
\begin{equation}
    dL = \pi dS \int_{\nu_1}^{\nu_2} \frac{2 h \nu^3}{\exp{\left(\frac{h\nu}{k_{\rm B}T_{\rm eff}}\right)} -1} d\nu
\end{equation}
where $\nu_1,\,\nu_2$ delimit the frequency interval. 
We note here that we do not model the emission coming from the corona that is supposed to form in these systems. In fact, the physics of the hot electrons constituting the corona is not captured by our simulations. Coronal emissions can be added in post processing assuming that a fraction of the circumBH discs luminosity is up-scattered by hot electrons, in analogy with standard AGNs. However, it is unclear whether such a population of diffused electron can efficiently build up in the presence of a cavity and possibly intermittent formation of circumBH discs in MBHB systems. We therefore decided here to conservatively omit this component from the emerging spectrum.

\subsection{Merger and post merger kicks}

We evolve our thin and thick disc simulations through the decoupling stage until merger. The two MBHs merge into one when they reach a separation of $6\,R_{\rm g}$, which corresponds to the Innermost Stable Circular Orbit (ISCO) for a Schwarzschild black hole.
After merger, we evolve the remnant only in the thin disc case. The newly formed sink particle has a radius of $6\,R_{\rm g}$ and we have simulated two kicks scenarios: one in the plane of the disc and one off plane, i.e. parallel to the disc angular momentum.
For the off-plane kick velocity we chose the average of the values reported in the last column of Table 2 in \cite{Baker2008}, i.e. ${\bf v}_{\rm kick}=(0,0,v_{z})$ with $v_{z}=530$ km/s. 
The in plane kick velocity vector is instead ${\bf v}_{\rm kick}= (v_0 \cos \phi, v_0\sin \phi, 0)$ with $v_0 = 94$ km/s and $\phi=\pi/2$, i.e. we assumed MBHs to have anti-aligned spins with magnitude $\chi =0.2$ and then used the expression for $v_{z}$ in Eq. 4 of \cite{Baker2008}.

\section{Results}
\label{sec:results}

In this section, we present our results and discuss the effects of the disc temperature and of the PN corrections on the dynamics of the MBHB. 
Furthermore, we discuss the scenario where the merger remnant experiences a kick in the plane of the disc and perpendicular to it. We will refer to the simulation with $H/R=0.03$ using the term ``cold disc'' and to the $H/R=0.1$ case with ``warm disc''. 

\begin{figure}
    \centering
    \includegraphics[width=\columnwidth]{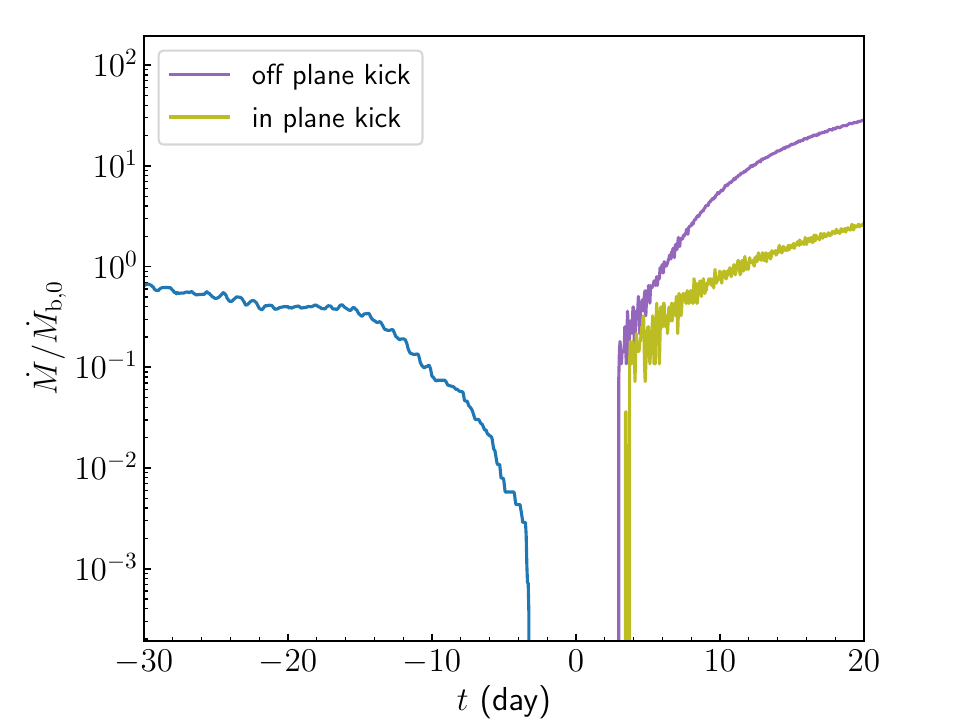}
    \caption{Accretion rate in units of the binary accretion rate at $t=0$ (corresponding to $a=100\,R_{\rm g}$) as a function of time. The blue curve represents the binary accretion rate until merger while the purple and green lines show the accretion rate of the MBH remnant following the off plane and in plane kick respectively.}
    \label{fig:mdot}
\end{figure}

\begin{figure*}
    \centering
    \includegraphics[width=0.49\textwidth]{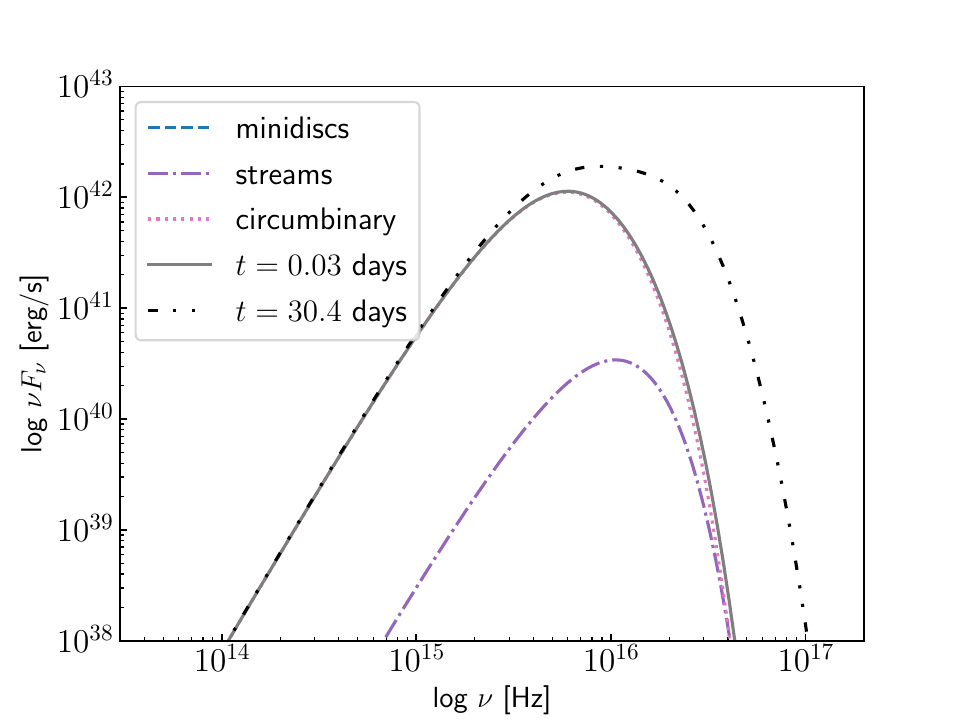}
    \includegraphics[width=0.49\textwidth]{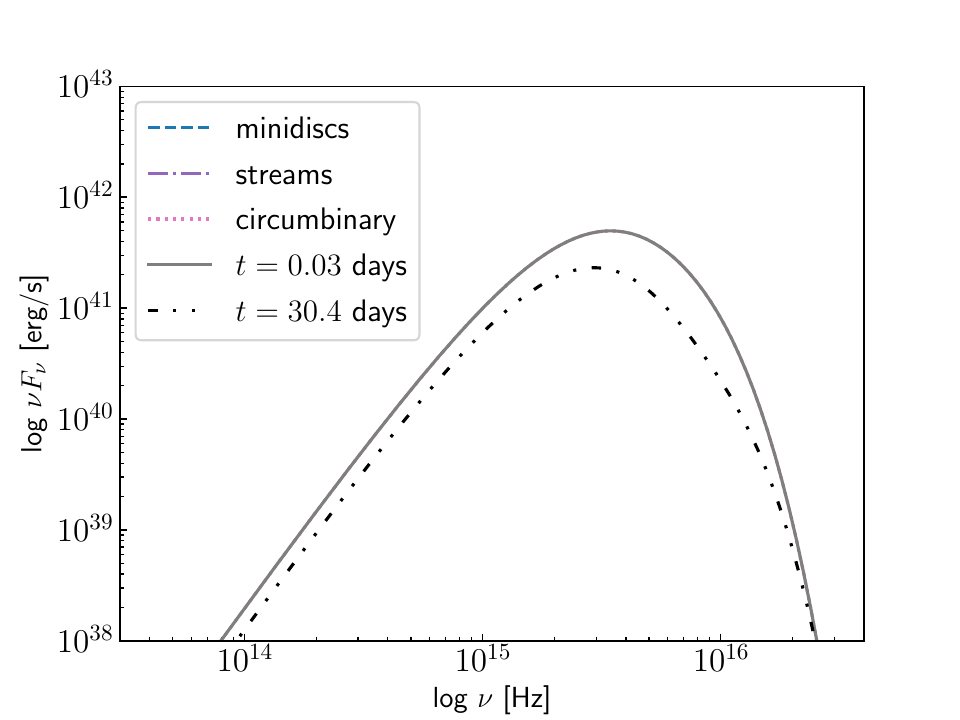}
     \caption{Spectrum of the black body emission from the warm (left panel) and cold (right panel) disc. The dot-dashed black line shows the SED $30.4$ days prior to merger for both simulations. The grey straight line shows the total SED right before merger. The blue dashed, purple dot-dashed and ppink dotted lines show the contribution from different regions of the disc $0.03$ days before merger. The blue dashed line shows the contribution of the disc inside $r=a$, which is absent in the pre-merger spectra while it contributes significantly to the X-ray flux of the initial (i.e. $t=30.4$ days) warm disc spectrum. The purple dot-dashed line shows the contribution from the gas at radii $a < r < 3a$ while the dotted pink line shows the circumbinary disc contribution.     } 
    \label{fig:spectra}
\end{figure*}

\begin{figure*}
    \centering
     \includegraphics[width=0.49\textwidth]{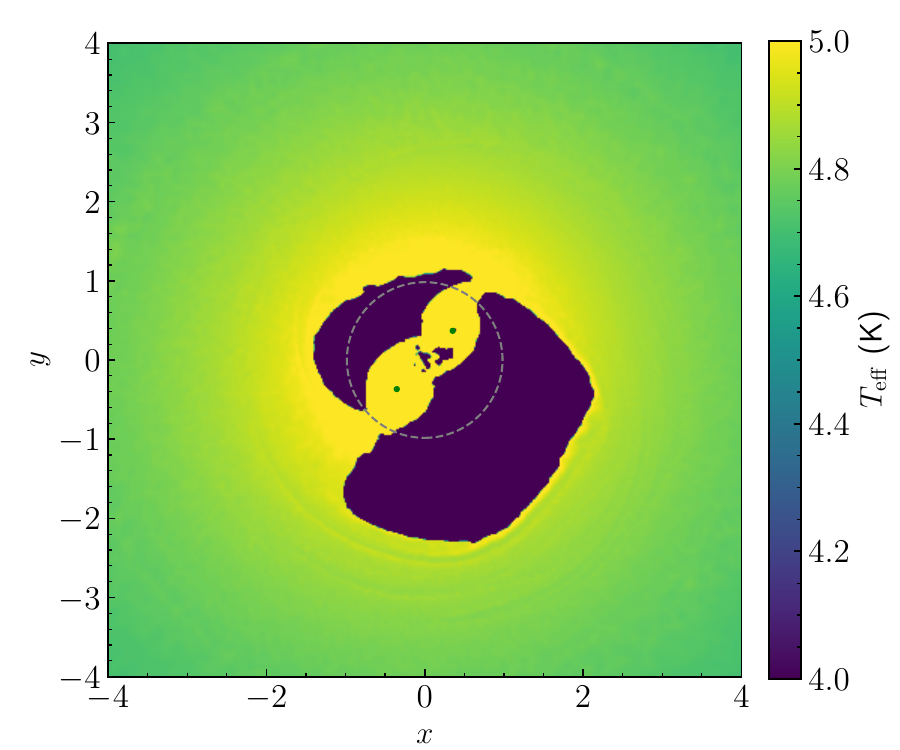}
    \includegraphics[width=0.49\textwidth]{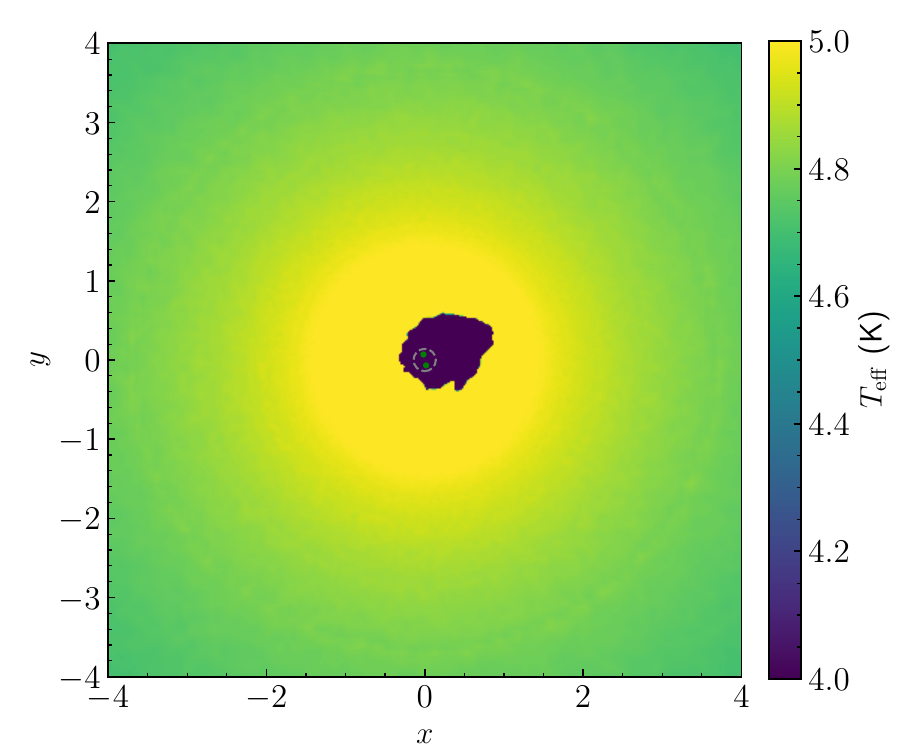}
    \includegraphics[width=0.49\textwidth]{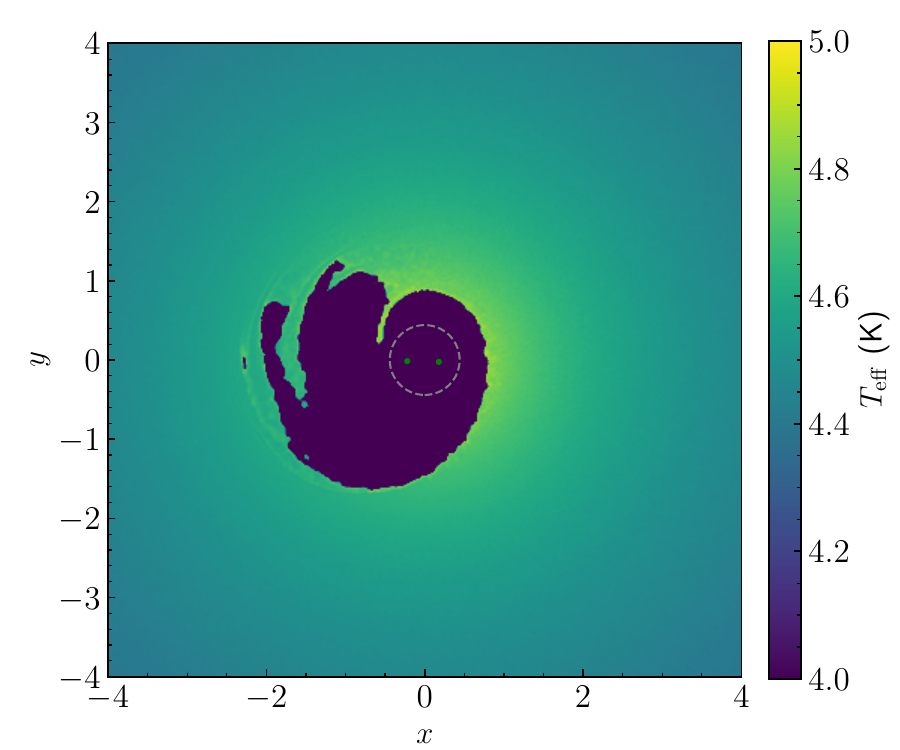}
    \includegraphics[width=0.49\textwidth]{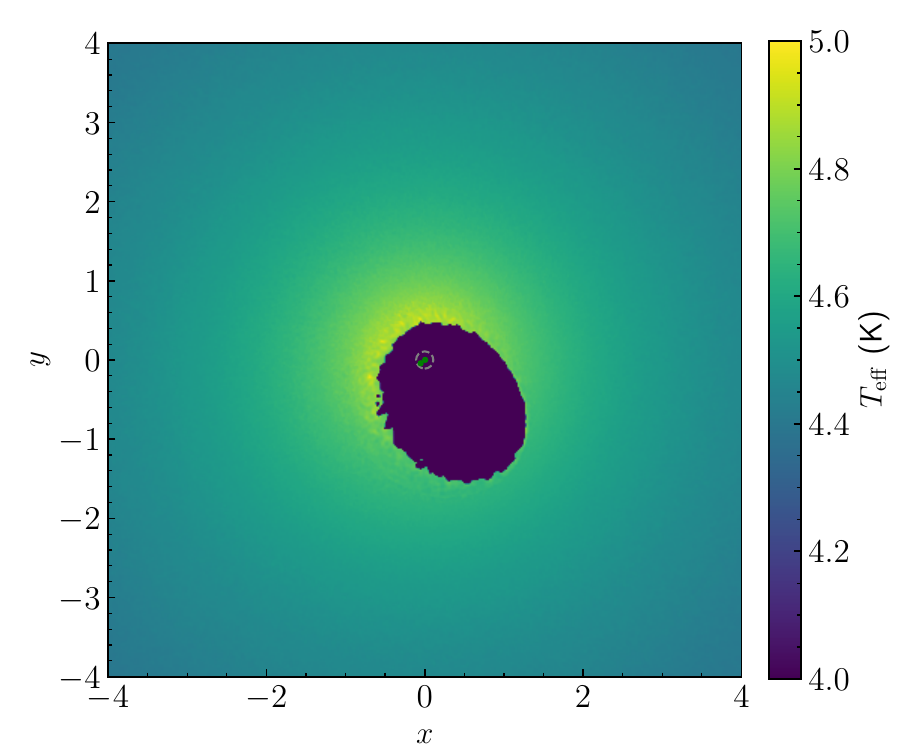}
    \caption{Maps of effective temperature $T_{\rm eff}$ for the warm (upper panels) and cold (bottom panels) disc simulation. The dashed circle encloses the binary orbit, following the binary inspiral.}
    \label{fig:teff}
\end{figure*}

\subsection{Warm disc}

A fairly large number of numerical studies has recently shown that warm discs (i.e. discs with aspect ratio $H/R\sim 0.1$) drive circular equal mass binaries to outspiral in the Newtonian regime \citep{Duffell2019,Munoz2020,tiede2020,heathnixon2020,Franchini2022}. 
GW emission at $\sim$100 $R_{\rm g}$ separation scales is going to dominate over any viscous torque, inevitably driving the binary to coalescence. Still, the presence of a positive net torque from a circumbinary disc can affect the dynamics by delaying the merger, thus introducing some modification to the phasing of the gravitational waveform."

We show the evolution of the binary semi-major axis in Figure \ref{fig:avst}.
The orange line shows the evolution of the binary in the purely Newtonian regime \citep{Franchini2022}, while the dot-dashed blue line shows the evolution obtained through the time integration of the PN terms only, i.e. neglecting the gas contribution.
The cyan line shows instead the evolution that we obtain in our simulation, which includes both the gas torque and PN corrections. The effect of the gas in this regime of disc aspect ratios is to drive the binary apart, effectively extending the inspiral phase. We note that the semi-major axis in the Newtonian simulation (orange line in Figure \ref{fig:avst}) expands to $1.001$ in $35$ days. However, with a disc mass as small as $100\,M_{\odot}$, we find this effect not to be strong enough to significantly delay the merger of the two MBHs. Note, however, that a tiny deviation from the evolution driven by the PN corrections can still be observed, as shown in the inset, as the cyan line slightly departs from the blue line. 

Even though the effect of gas on the binary semi-major axis evolution is not significant in this case, there might be a non-negligible effect in terms of the orbital phase of the binary.
We traced the evolution of the orbital phase over time in the warm disc simulation to infer whether the presence of the gaseous disc induces identifiable phase shifts with respect to the PN only evolution. The results are shown in the upper panel of Figure \ref{fig:phase} where the cyan and dot dashed line show the phase in the simulation with and without the gaseous disc respectively. 
Although the effect is small, the presence of a warm disc causes a small delay in the coalescence of the binary. As an illustrative example, we consider the delay accumulated starting from five days prior to merger. At this point, an equal mass binary of total mass $10^6$M$_\odot$ is emitting at a frequency of $\approx1.6\times10^{-4}$Hz, and starts to be resolvable by LISA. Within these last five days, the torques from the warm disc cause a delay of the merger of about 10 seconds. Considering $f_{\rm ISCO}\approx4\times10^{-3}$Hz for our binary, this implies a shift of about 10 seconds over a period of 250 seconds at ISCO, which corresponds to 0.04 cycles or a phase shift of 0.25\,rad.

Figure \ref{fig:warm} shows column density maps of the circumbinary disc surrounding the MBHB starting from $30.4$ days (upper left panel) prior to merger all the way down to the instant right before merger (lower right panel).
The decoupling should occur at around 10 days but we find the gaseous disc to be able to feed material to the binary down to $\sim 1$ day from the merger.
We can clearly see from the last panel of Figure \ref{fig:warm} that the discs around the binary components disappear within the last day prior to the merger, as the disc is not able to follow the inspiral any longer.
We can therefore possibly expect some sort of EM precursor, at least in the case of warm discs, as the X-ray emission coming from the material inside $r=a$ disappears.

\subsection{Cold disc}

Previous numerical investigations of the interaction between a MBHB and a cold disc have shown that the disc torque is negative and therefore contributes to the inspiral of the binary \citep{tiede2020,heathnixon2020,Franchini2022}. 
We therefore expect the presence of the circumbinary disc to speed-up the coalescence process compared to pure PN dynamics in this case. 

The binary takes about 450 days to coalesce from an initial separation of 120$R_{\rm g}$ (bottom panel inset of Figure \ref{fig:avst}). The difference in the merger timescale owing to the presence of the gas, computed comparing the cold disc simulation to a simple integration of the binary orbit without the gaseous disc, is around $0.02\%$, meaning that the presence of the gas brings the two MBHs to coalescence about 2 hours  earlier compared to pure PN dynamics. Thus, the presence of the gas has a more significant impact on the evolution of the system with respect to the warm disc case. In particular, the lower panel of Figure \ref{fig:phase} shows that the disc torques anticipate the merger by about 130 seconds in the last 5 days prior to the merger. At ISCO, this corresponds to a difference of about half a cycle or $\pi\,$rad in the phasing of the waveform. 

Figure \ref{fig:cold} shows the column density maps of the circumbinary disc surrounding the MBHB starting from 30.4 days prior to merger, which corresponds to a phase where the binary should have already decoupled from the gaseous disc.
The last panel shows the disc right before merger. The over-density, or "lump" \citep{Shi2012,farris2014,ragusa2016}, persists down to the very last stages of the merger, even though the gaseous disc is expected to have decoupled from the binary.

The circumBH discs are not present as in this regime of disc aspect ratios there is not enough material that flows inside the cavity to be able to resolve the discs structures in the case of constant $\alpha$ discs, as the ones that are treated in this work. 


\subsection{Post merger kicks}

We here investigate the accretion onto the merger remnant, considering that it may be subject to post-merger kicks \citep{Baker2008,Rossi2010}.
We present the results of the in plane and off plane kicks in the cold disc simulation only as this is possibly the most interesting case due to the highly eccentric cavity.

In the case of an in-plane kick along the $y$ axis, the remnant moves towards the cavity pericentre, i.e. upwards with respect to the plane centre in the bottom right panel of Figure \ref{fig:cold}. We find the cavity to become more circular and progressively smaller, although on a slower timescale compared to the motion of the remnant. 
The kick along the $z$ axis causes instead the disc to fill the cavity on a slightly shorter timescale and then to follow the remnant dynamics, starting from the inner parts.

Figure \ref{fig:mdot} shows the accretion rate, 
traced over the entire duration of the cold disc simulation, for both the in-plane (green line) and off-plane (purple line) kick scenarios. This is scaled by the initial accretion rate in the Newtonian simulation. However when scaled to physical values, the accretion rate is $\sim 10\,\dot{M}_{\rm Edd}$  and $\sim 10^3\,\dot{M}_{\rm Edd}$ in the cold and warm disc simulations respectively.
The accretion rate in the off-plane kick case increases more rapidly as, once the remnant starts moving rapidly along the $z$-axis, the inner parts of the disc quickly follow its off-plane motion.
The in-plane kick, however, is along the $y$-axis and has a smaller velocity. The merger remnant then moves slowly towards the cavity pericentre (see bottom right panel of Fig. \ref{fig:cold}) and accretes at a much slower rate because of the cavity remaining eccentric and wider for longer. Note that we find the cavity eccentricity to increase with time, reaching its peak ($e_{\rm disc}=0.6$) at the merger. 

Furthermore, we can clearly see from Figure \ref{fig:mdot} that the binary accretion rate drops by several orders of magnitude at merger (as expected) to increase again by several orders of magnitude after the merger, the slope depending on whether the kick experienced by the remnant is in plane or perpendicular to the disc plane.
We find that there is a $\sim 1$ day delay in the onset of the accretion onto the remnant between the off plane and in plane kick case.

\subsection{Electromagnetic signatures}

In this section, we discuss the possible EM signatures of MBHBs before their merger. 

\subsubsection{Spectral Energy Distribution}

We show the SEDs obtained from the warm and cold disc simulations in the left and right panel of Figure \ref{fig:spectra} respectively.
The dot-dashed black line shows the SED $30.4$ days before merger for the warm and cold disc simulation while the grey solid lines show the total SED at the last stage before merger (i.e. bottom right panel of Figures \ref{fig:warm} and \ref{fig:cold}). 
The blue dashed, purple dot-dashed and pink dotted lines show the contribution to the SED from radii $r<a$, $a<r<3a$ and $r>3a$ respectively right before merger. Note that the circumBH discs contribution (i.e. blue dashed line) is absent in both cases as prior to merger there is no material close enough to the binary to emit in the X-rays.

We show the effective temperature we used to compute the SEDs, $30.4$ days prior to merger and $0.03$ days before merger in Figure \ref{fig:teff}. The upper panels refer to the warm disc case while the lower panels show the temperature map in the cold disc case. 
These maps reflect the characteristics of the spectra shown in Figure \ref{fig:spectra}.

\begin{figure*}
    \centering
    \includegraphics[width=0.49\textwidth]{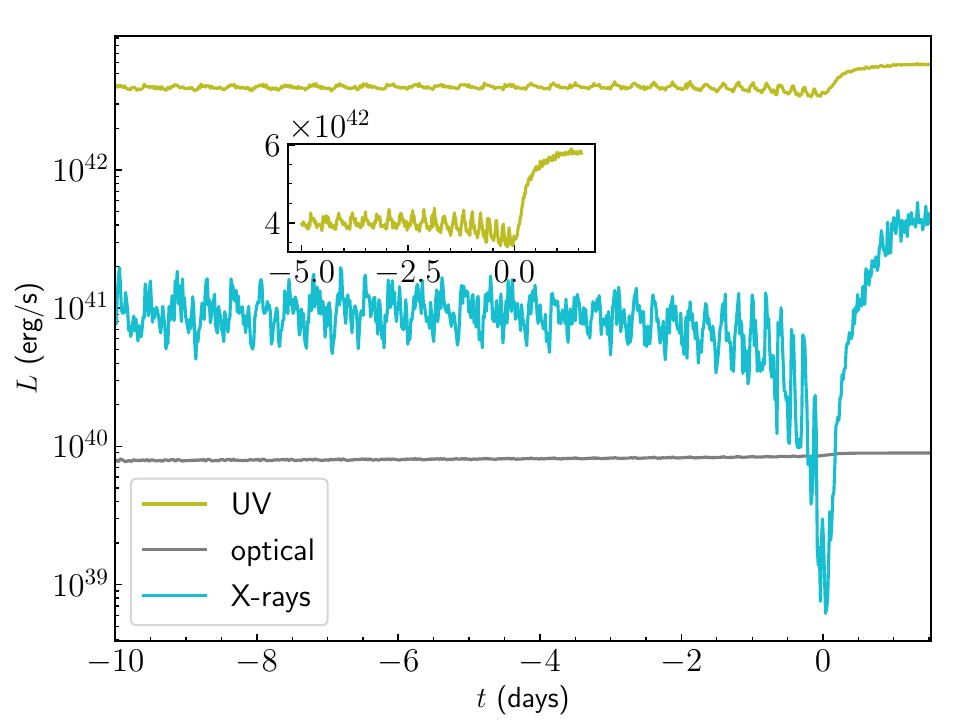}
    \includegraphics[width=0.49\textwidth]{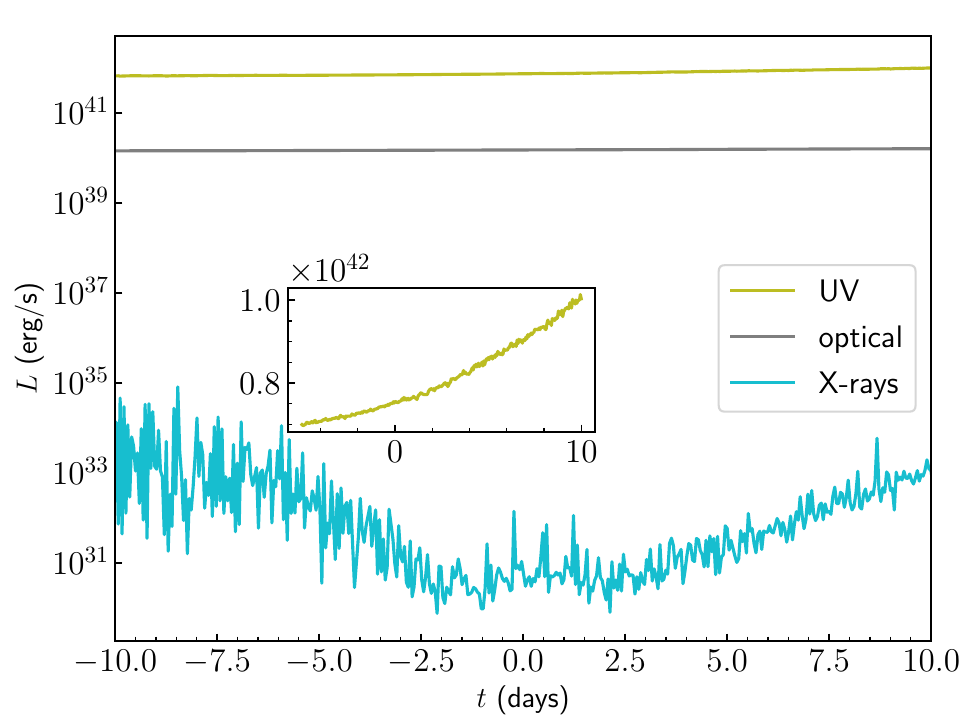}
    \caption{Fraction of the total luminosity emitted by the whole simulation domain in different bands, for the warm (upper) and cold (lower) disc simulation. }
    \label{fig:L}
\end{figure*}

In the warm disc case (left panel of Fig. \ref{fig:spectra}), the emission becomes dominated by the gas outside $r=3a$ within the last $0.03$ days as the circumBH discs are accreted onto the binary components and the circumbinary disc has decoupled from the binary without being able to supply more material into the cavity. 
The pre-merger spectrum therefore peaks in the optical/UV and shows no X-ray tail.
We can indeed see from the upper right panel of Figure \ref{fig:teff} that the X-ray emission region is significantly reduced right before merger.

In the cold disc case (right panel of Fig. \ref{fig:spectra}) the cavity remains very eccentric even at the merger.
As the cavity closes around the binary, the over-density at the cavity edge becomes warmer leading to an increase in the UV flux by a factor 2 in 10 days following the merger. The optical flux remains roughly constant.

The bottom right panel of Figure \ref{fig:teff} shows that the slight increase in the UV flux in the spectrum (i.e. grey line in the right panel of Fig. \ref{fig:spectra}) is caused by a slightly more extended warm region at the cavity edge right before merger, as the cavity is closing around the coalescing binary. 

\subsubsection{Emission at different wavelengths}

Since the variations in the SED of these systems might be potentially difficult to detect, it is important to investigate the EM emission at different wavelength and how this changes approaching the merger.
In particular, as the gaseous disc decouples from the binary we expect the high energy emission to decrease as the spectrum peak shifts towards lower frequencies.

Figure \ref{fig:L} shows the luminosity emitted in each frequency band: optical ($1.8-3.1$ eV), UV ($3.1-124$ eV) and X-rays ($124$ eV $-124$ keV) for the warm (left panel) and cold (right panel) disc simulations as a function of time. The negative times refer to the pre-merger evolution.

The first clear difference between the warm and cold disc cases, i.e. between left and right panel of Figure \ref{fig:L} respectively, is that initially (i.e. 30.4 days prior to merger) the flux in the X-ray band is more prominent in the warm disc case while in the cold disc simulation is essentially negligible. The reason for such low values of X-ray fluxes in the cold disc case is that the cavity optical depth is extremely small and we therefore cannot apply our black body approximation to compute the emitted flux.
The UV emission is however predominant in both cases, which is consistent with the spectra in Figure \ref{fig:spectra} that shows the peak to be shifted towards UV/optical frequencies.
This emission increases by roughly a factor 2 in both cases but on a shorter timescale, i.e., within 2 days following the merger, in the warm disc case.
This difference is likely due to the fact that the cavity in the warm disc case is significantly less eccentric and smaller, therefore closer to the merger remnant. In the cold disc case instead the cavity is very eccentric and takes a longer time to close around the merger remnant, therefore delaying the rise in the UV flux.

In the warm disc case we can see that roughly one day before merger the X-ray flux drops by at least two orders of magnitude, consistently with the drop in the accretion rate onto the binary, while the increase in UV and optical flux is milder. 
The optical flux remains roughly constant and significantly subdominant compared to the UV emission. 

The cold disc case shows a different behaviour, where the UV flux increases at merger, together with the X-ray flux, see right panel of Figure \ref{fig:L}, as the cavity closes around the remnant. 
We note that we have tested the dependence of our results on the resolution inside the cavity in this case, increasing the refinement by a factor 4 in order to better resolve the emission from the circumBH discs.
We find the X-ray flux to change by at most an order of magnitude, therefore remaining much lower than the UV/optical flux. 
We therefore caution that the magnitude of the X-ray flux is sensitive to our resolution inside the cavity. We can however capture variations in the X-ray flux, which is ultimately one of the possible EM imprints of the presence of a MBHB, regardless of the resolution inside the cavity.

\begin{figure}
    \centering
    \includegraphics[width=\columnwidth]{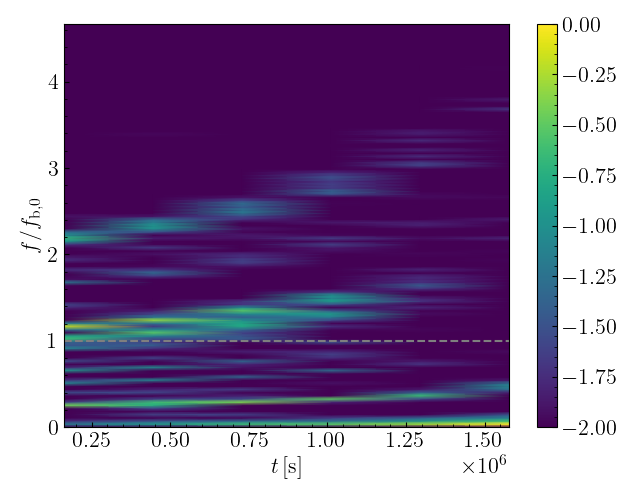}
    \caption{Spectrogram of the flux in the optical band for the warm disc simulation. The dashed grey line marks the initial binary orbital frequency.}
    \label{fig:modulation}
\end{figure}

Finally, we identified two periodic modulations of the flux: one on the orbital period of the binary and the other on the period associated with the motion of the overdensity located at the cavity edge, i.e. around 5 times the orbital period. These are both expected in equal mass binaries fed from a circumbinary disc \citep{Duffell2019,Munoz2019}. 
We find these modulations to be more evident in the UV and X-rays bands. However we find that they can also be identified in the optical band, which is targeted by the current ZTF \citep{ZTF} survey and by the upcoming Vera Rubin Observatory \citep{VRO}.
Figure \ref{fig:modulation} shows the spectrogram of the optical flux for the warm disc simulation.
The binary orbital period modulation is slightly stronger in the UV and X-rays band than in the optical.
The overdensity modulation is also stronger in the UV and in the X-rays than in the optical. 
We however show the optical periodogram as this is the most interesting frequency band for current surveys such as ZTF and the Rubin Observatory.
Note that the strength of the modulation on the orbital period decreases with time towards the merger as our time sampling rate becomes too sparse to resolve the binary orbital period.
The cold disc simulation shows similar features that do not evolve with time as the cavity is further away from the binary orbit and therefore the gas motion is not strongly influenced by its inspiral.

\section{Conclusions}
\label{sec:conclusions}

In this work, we have investigated the dynamics and emission properties of a coalescing $10^6\,M_{\odot}$ MBHB embedded in a $100\,M_{\odot}$ circumbinary disc following the inspiral with up to 2.5PN correction to the binary dynamics. We have run two numerical simulations with different disc temperatures (i.e. different disc aspect ratio) in order to understand the type of emission that we could expect to detect in the two cases. 
We used the 3D hydrodynamics code {\sc gizmo} in MFM mode with hyper-Largangian refinement in order to properly resolve the circumBH discs and therefore their emission. We also follow the dynamics of the MBH remnant after merger and investigate the effect that an off-plane and in-plane kicks have on the accretion rate.

We compared the evolution of the binary semi-major axis in both simulations to the evolution due to the sole PN terms, without the gas contribution. We find the difference to be negligible in the warm disc case while in the cold disc case the binary merges on a timescale shorter by $\sim 2$ hours in the presence of the gaseous cold circumbinary disc.

We also compared the evolution of the orbital phase of the binary in the warm and cold disc simulation, finding a phase shift, caused by the presence of the gaseous disc, of about 0.25 rad and $\pi$ rad respectively (see Fig. \ref{fig:phase}). 
We might be able to obtain a more precise estimate of the observability of this phase shift by calculating the useful cycles, which is a signal to noise ratio dependent quantity \citep{Sampson2014}.
The detailed investigation of the detectability of this effect is the subject of a future work, but our simulation suggests that the presence of a dense, cold circumbinary disc might leave appreciable imprints in the GW waveform also beyond decoupling.

We computed SEDs from the optically thick regions of the domain. 
We then integrated the flux over the whole radial domain and obtained the light curves in different bands: optical, UV and X-rays. 
We find the shape of the SED to change significantly towards the merger, with the X-ray flux disappearing in the warm disc case and the UV flux moderately rising in the cold disc case. 
Consistently, we find the same behaviour in the light curves, where the decrease in the X-ray flux at the merger by two orders of magnitude is accompanied by a factor 2 increase in the UV flux. 
The decrease in the X-ray flux is likely associated with the disappearing of the circumBH discs. 
We note that the drop in the X-ray flux we find is similar to the one reported by \cite{MajorKrauth2023} and \cite{Dittmann2023}. Since we do continue to evolve the system after the merger, the cavity closes around the merger remnant quite rapidly as the disc is viscous, therefore restoring the X-ray emission.
The cold disc case shows instead a very mild variation in the subdominant X-ray flux and a corresponding increase by a factor 2 in the UV flux after the merger. In this case the rise in UV flux is slower compared to the warm disc case, and is due to the cavity being wider and the disc being colder, hence less viscous, which reflect in a longer time needed to fill the cavity around the remnant.
We also note that, in the warm disc case, our initial spectrum peaks at a lower frequency compared to the spectrum presented in \cite{2018MNRAS.476.2249T}. This is possibly due to the different choice of the equation of state, i.e. locally isothermal (this work) or adiabatic \citep{2018MNRAS.476.2249T}. However, pinpointing the cause for this discrepancy is not straightforward as our simulations are 3D and therefore not directly comparable with 2D runs (Duffell et al. in prep).

Finally, we looked into periodicities in the UV, optical and X-ray flux, finding two characteristic modulations in all bands: one on the binary orbital period and one on the timescale of the precession of the overdensity at the cavity edge.
The first is more important in the warm disc simulation as there is significantly more material in the circumBH discs with respect to the cold disc simulation. The latter, super-orbital modulation, is instead slightly more prominent in the cold disc case.
Even though these modulations are more easily detectable in the high-energy bands (UV and X-rays), they also appear in the optical band, making it possible for surveys such as ZTF and the Legacy Survey of Space and Time (with the Vera Rubin Observatory) to identify MBHB candidates prior to their merger.

\begin{acknowledgements}

AF and AS acknowledge financial support provided under the European Union’s H2020 ERC Consolidator Grant ``Binary Massive Black Hole Astrophysics" (B Massive, Grant Agreement: 818691). MB acknowledge support provided by MUR under grant ``PNRR - M4C2 - I1.2  - Avviso 247 del 19.08.2022 e ss.mm.ii, ID:SOE\_0163''.
\end{acknowledgements}

%
%
\bibliographystyle{aa} 
\bibliography{biblio,references}

\end{document}